\documentclass{PoS}

\usepackage{graphicx}
\usepackage{graphics}
\usepackage{tikz}
\usepackage{makecell}
\usepackage[mathscr]{eucal}
\usepackage{amssymb}
\usepackage{amsmath}
\usepackage{xspace}
\usepackage{listings}
\usepackage{ulem}
\usepackage{color}
\usepackage{alltt}
\usepackage[percent]{overpic}
\usepackage{subfigure}

\newcommand{\BL}{{\ensuremath{B-L}}\xspace}
\newcommand{\UBL}{{\ensuremath{U(1)_{B-L}}}\xspace}
\newcommand{\vev}[0]{VEV\xspace}
\newcommand{\vevs}[0]{VEVs\xspace}

\def\gsim{\raise0.3ex\hbox{$\;>$\kern-0.75em\raise-1.1ex\hbox{$\sim\;$}}}

\newcommand{\vevacious}[0]{\texttt{Vevacious}\xspace}
\newcommand{\vcs}[0]{\vevacious}

\definecolor{darkgreen}{rgb}{0,0.5,0}

\newcommand{\BLSSM}{BLSSM\xspace}

\newcommand{\rsnu}{R-sneutrino\xspace}
\newcommand{\rsnus}{R-sneutrinos\xspace}

\newcommand{\TAB}{Tab.\xspace}

\newcommand{\RP}[0]{$R$ parity\xspace}

\newcommand{\RPVg}[0]{\RP-violating\xspace}

\newcommand{\RPCg}[0]{\RP-conserving\xspace}
\newcommand{\SUL}{{\ensuremath{SU(2)_{L}}}\xspace}

\title{On the vacuum stability of SUSY models}

\ShortTitle{On the vacuum stability of SUSY models}

\author{\speaker{Jos\'{e} Eliel Camargo-Molina},$^a$ Ben O'Leary,$^a$ Werner Porod $^a$ and Florian Staub $^b$ \\
\llap{$^a$} Institut f\"ur Theoretische Physik und Astronomie,
Universit\"at W\"urzburg\\
Am Hubland,
97074 W\"urzburg, Germany \\
\llap{$^b$} Bethe Center for Theoretical Physics \& Physikalisches Institut der
 Universit\"at Bonn, \\
53115 Bonn, Germany \\

E-mail: \email{jose.camargo@physik.uni-wuerzburg.de}, \email{ben.oleary@physik.uni-wuerzburg.de}, \email{porod@physik.uni-wuerzburg.de}, \email{fnstaub@th.physik.uni-bonn.de}

}

\abstract{The existence of multiple non-equivalent minima of the scalar potential in SUSY models both raises technical challenges and introduces interesting physics. The technical challenges are now that one has to find several minima and evaluate which is the deepest, as well as calculate the tunneling time from a false vacuum to the true vacuum. We present here studies on the vacuum stability and color/charge breaking minima in the CMSSM and $R$ parity violating minima in a B-L extended MSSM.}

\FullConference{The European Physical Society Conference on High Energy Physics \\
		 18-24 July, 2013\\
		 Stockholm, Sweden}

\begin{document}

\section{Introduction}
An important part of the phenomenology of the incredibly successful standard model of
 particle physics (SM) is the spontaneous breaking of some (but not all) of the
 gauge symmetries of the Lagrangian density by the
 \textit{vacuum expectation value} (\vev) of a scalar field charged under a 
 subgroup of the SM gauge group. The entire scalar sector
 of the SM consists of a doublet of $SU(2)_{L}$ with $-\frac{1}{2}$ $U(1)_{Y}$ hypercharge. 
 The potential energy of the vacuum is
 minimized by the scalar field taking a constant non-zero value everywhere. The
 presence of this \vev 
 allows for massive particles that would be forced to be massless if the
 gauge symmetries of the Lagrangian density were also symmetries of the vacuum
 state.
 
Many extensions of the SM introduce extra scalar fields. Sometimes these fields
 are introduced explicitly to spontaneously break an extended gauge symmetry
 down to the SM gauge group e.g.\cite{Langacker:2008yv, Basso:2010si}, and they are
 assumed to have non-zero
 \vevs at the true vacuum of the theory. Other times they are introduced for
 other reasons, such as supersymmetry \cite{Nilles:1983ge}, and
 often non-zero \vevs for
 such fields would be disastrous, such as breaking $SU(3)_{c}$ and/or
 $U(1)_{\text{EM}}$, which excludes
 certain parts of the parameter space of the minimal
 supersymmetric standard model (MSSM) from being phenomenologically relevant. The existence 
 of multiple non-equivalent vacua both raises technical challenges
 and introduces interesting physics. The technical challenges are now that one
 has to find several minima and evaluate which is the deepest, as well as
 calculate the tunneling time from a false vacuum to the true vacuum. 

The technical challenges are much tougher when multiple scalar fields are
 involved. Even a tree-level analysis involves solving a set of coupled cubic
 equations, the so-called minimization or tadpole equations. It has generally
 only been attempted for highly symmetric systems such as two Higgs doublet
 models \cite{Lee:1973iz, Branco:2011iw} or with only a minimal amount of
 extra degrees of freedom such as the (assumed) three non-zero \vevs of the
 next-to-minimal supersymmetric standard model \cite{Fayet:1974pd, Ellis:1988er, Drees:1988fc}.

The program \vcs \cite{CamargoMolina:2013qva} has been written to address this.
 Given a set of tadpole
 equations and the terms needed to construct the one-loop effective
 potential,
\vcs brings together a series of public tools to find the minima of one-loop
potentials and if required calculate the tunneling time between them. 


We present here highlights from studies performed 
on the vacuum stability of two supersymmetric models.  In section 2 we showcase some results from \cite{Camargo-Molina:2013sta}, 
where we study the
color and charge breaking minima that might appear in the mSUGRA inspired
Constrained MSSM (CMSSM). In section 3 we discuss some results from \cite{CamargoMolina:2012hv}, where we investigate 
the issue of $R$ parity violation in the B-L extended CMSSM, as for this model $R$ parity can be violated
spontaneously through \vevs for the scalar partners of right-handed neutrinos.  

 \section{Revisiting the vacuum structure of the CMSSM}
\begin{figure}[!htbp]
\begin{center}
\includegraphics[width=0.65\linewidth
               ]{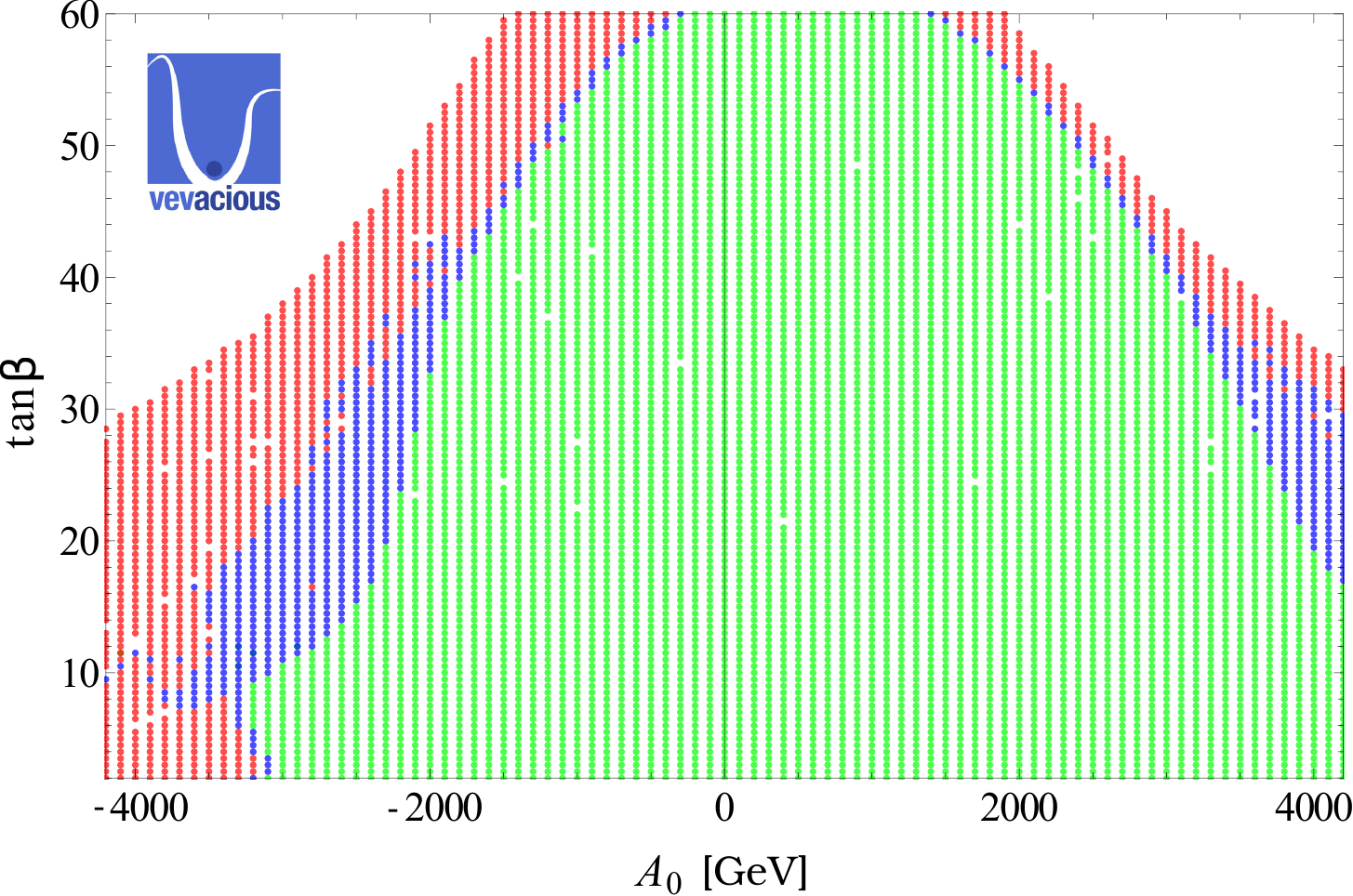}  

\caption{Projection in the $\tan\beta$ / $A_0$ plane for a random scan in the CMSSM ($M_0$ and $M_{1/2}$ are fixed at 1 TeV). Green points have the input color- and charge- conserving minimum as global minimum. Blue and red points
have charge- and color- breaking minima which are deeper than the input minimum. Red points have tunneling times from the
input minimum to the CCB minimum lower than one tenth of
 the age of the Universe. Blue points have a tunneling time greater than a
 tenth of the age of the Universe.}
\label{fig:A0_tanb_prelim}
\end{center}
\end{figure}
\subsection{The model}
One popular way of ameliorating the standard model hierarchy problem is to promote it to
 a supersymmetric theory, such as the MSSM \cite{Martin:1997ns}. The mechanism of supersymmetry (SUSY)
 breaking introduces
 more parameters and the breaking of SUSY is often parametrized by adding
soft SUSY-breaking terms to the Lagrangian density. The number of
 parameters specifying the full set of soft SUSY-breaking terms allowed in the
 MSSM is rather large, namely 105, so often
 they are taken to be related at a specific scale. One of the
 simplest and most
 popular proposal is the minimal-supergravity-inspired
 constrained MSSM (CMSSM), in which all the soft SUSY-breaking scalar
 mass-squared terms and the mass terms
for the fermionic partners of the gauge bosons are taken to be equal to $M_{0}^{2}$ and $M_{1/2}$ respectively at the scale 
($M_{GUT}$) where the gauge couplings
 unify. The $A$ terms, which multiplied by the
 corresponding Yukawa couplings give the trilinear scalar interaction couplings, are also taken to unify at the value $A_0$. 
 $|\mu|$ and $B_{\mu}$ are fixed by requiring that the mass of the $Z$ boson is
 correct along with defining the ratio $\tan \beta$ of the
\vevs $v_{d}$ and
 $v_{u}$ of the neutral components of the two Higgs doublets, and the sign of
 $\mu$ is given as a final input.

The presence of many additional scalar
 partners for the SM fermions raises the question of whether they too could
 develop \vevs. Unfortunately, until recently
 it was quite impractical to search for other vacua to see whether the desired
 vacuum is stable, or whether there are charge- and/or color-breaking (CCB)
 minima.

 We use \vcs to
 investigate regions of the CMSSM which, despite having local minima with the
 desired breaking of $SU(2)_{L} \times U(1)_{Y}$ to $U(1)_{EM}$ while preserving
 $SU(3)_{c}$, might have global minima with a different breaking of the gauge
 symmetries.

\subsection{Results}

In fig.\ref{fig:A0_tanb_prelim} we present results 
from a random scan in the $A_0$ / $\tan\beta$ plane where the GUT-scale parameters $M_0$ and $M_{1/2}$ were fixed at 1 TeV and $\mu$ was taken to be positive.  

In this particular plot it is interesting to see that a large ``green area'' remains where the input minimum is the global one. However, a considerable amount of points develop stop and stau \vevs in areas of the parameter space that are not otherwise excluded by experiment. This shows the potential use of vacuum stability and the appearance of color- and charge- breaking minima as a phenomenological constraint. It is worth noting that for $\tan \beta < 10$  we encountered points with mostly stop \vevs reflecting that enhancement of $Y_{t}$ might play a significant role, whether for high $\tan \beta$ stau \vevs were dominant as now $Y_{\tau}$ and $A_{\tau}$ are significantly enhanced, thus making it easier for stau \vevs to develop. It is interesting to point out that to obtain an LSP below 1 TeV (important for Dark Matter studies), light staus are needed. This in turn implies large values of $|A_0|$ and/or $\tan\beta$. As can be seen in fig.\ref{fig:A0_tanb_prelim}, this is a dangerous area as very often we get stau \vevs.

\section{Stability of $R$ parity in the BLSSM}

\subsection{The model}

There are several ways to extend the MSSM by \UBL. For this study, we confine ourselves to the minimal of such extensions, 
 which allows for a spontaneously broken \UBL without necessarily
 breaking \RP. This requires the addition of two SM gauge-singlet chiral
 superfields $( \eta, \tilde{\eta})$ carrying \BL charge which have to develop \vevs for reasonable phenomenological results, as well as the addition of
 three generations of superfields containing right-handed neutrinos. Analogous to the CMSSM, we reduce the number of free
parameters by assuming unification of the gaugino masses, the soft SUSY-breaking scalar squared masses and trilinear couplings. We refer to
 this model as the \BLSSM.

The \BLSSM has a rich phenomenology, with a  $Z'$ boson  \cite{Krauss:2012ku}, Majorana neutrinos with see-saw
 masses, several qualitatively new dark matter candidates  \cite{Basso:2012gz} and a rich Higgs sector  \cite{Basso:2012tr}. However most of the  phenomenological studies for the \BLSSM  assume that \RP is conserved.  
 
 The main idea behind investigating the vacuum structure of the \BLSSM is to get a picture of how
 robustly \RP is conserved for parameters of phenomenological interest, as \UBL might be spontaneously broken by \vevs for the scalar partners of the right handed neutrinos (\rsnus) instead of the being broken by $\eta$ and $\tilde{\eta}$ fields thus generating \RPVg interactions. 
 \begin{figure}[!htbp]
$\begin{array}{cc}
\begin{overpic}[scale=0.55]{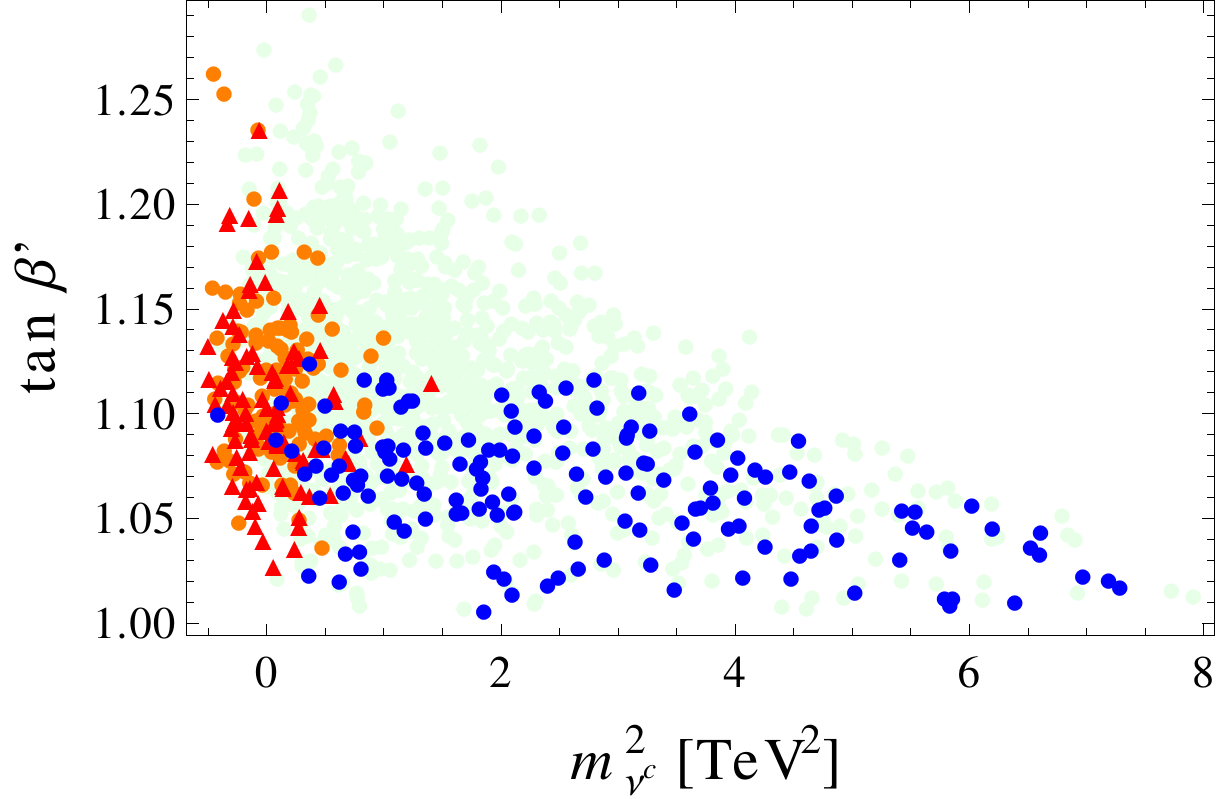}
     \put(80,45){\includegraphics[scale=0.075]{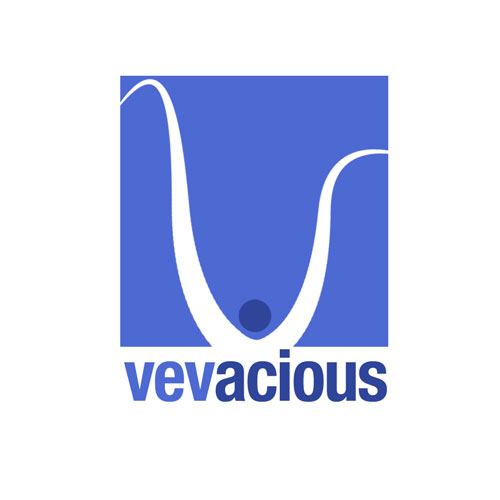}}  
  \end{overpic} \ \ \ \ \ 
&

\begin{overpic}[scale=0.56]{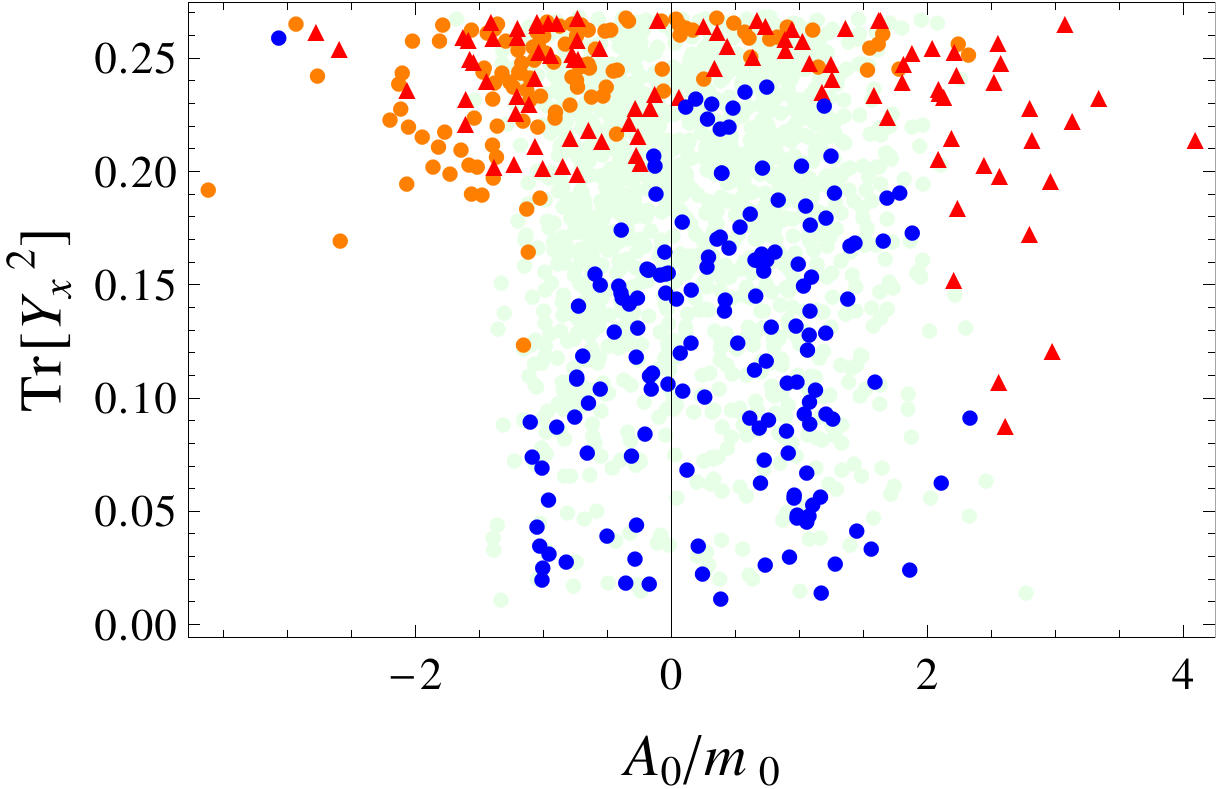}
     \put(17,15){\includegraphics[scale=0.075]{vevaciouslogo.jpg}}  
  \end{overpic}
\end{array}$

\caption{Projections onto various parameter planes of the 1640 hierarchical scan
 parameter points, categorized by the nature of their global minima at the one-loop level. $m^2_{\nu^c}$ denotes the smallest soft SUSY-breaking mass-squared for sneutrinos.
 ``RPC'' points are plotted in green, ``RPV'' points
 are in two groups based on the tunneling time from the
 ``RPC'' input minimum to the deeper ``RPV'' minimum: lower than one tenth of
 the age of the Universe as orange circles, greater than a
 tenth of the age of the Universe as red triangles.
 ``Gauge conserving'' points are in blue.}
\label{fig:hier_loop_comparisons}
\end{figure}

$\eta$ is responsible for
generating a Majorana mass term for the right-handed neutrinos and
thus we interpret the \BL charge of this field as its lepton number,
and likewise for $\bar{\eta}$, and call these fields bileptons since
they carry twice the lepton number of (anti-)neutrinos.  We summarize
the quantum numbers of the extra chiral superfields with respect to the
 model's gauge group $U(1)_Y \times SU(2)_L \times SU(3)_C \times \UBL$ in
 \TAB~\ref{tab:cSF}.
\begin{table} [!htbp]
\centering
\begin{tabular}{|c|c|c|c|c|c|} 
\hline \hline 
Superfield & Spin 0 & Spin \(\frac{1}{2}\) & Generations & \, \( U(1)_Y\otimes\,
SU(2)_L\otimes\, SU(3)_C\otimes\, \UBL \) \, \\ 
\hline 

\(\hat{\nu}^c\) & \(\tilde{\nu}^c\) & \(\nu^c\) & 3
 & \((0,{\bf 1},{\bf 1},\frac{1}{2}) \) \\ 
\(\hat{\eta}\) & \(\eta\) & \(\tilde{\eta}\) & 1
 & \((0,{\bf 1},{\bf 1},-1) \) \\ 
\(\hat{\bar{\eta}}\) & \(\bar{\eta}\) & \(\tilde{\bar{\eta}}\) & 1
 & \((0,{\bf 1},{\bf 1},1) \) \\ 
\hline \hline
\end{tabular} 
\caption{Extra chiral superfields, in addition to those of the MSSM, and their quantum numbers.}
\label{tab:cSF}
\end{table}

The superpotential is given by 
\begin{align} 
\nonumber 
W = & \, Y^{ij}_u\,\hat{u}^c_i\,\hat{Q}_j\,\hat{H}_u\,
- Y_d^{ij} \,\hat{d}^c_i\,\hat{Q}_j\,\hat{H}_d\,
- Y^{ij}_e \,\hat{e}^c_i\,\hat{L}_j\,\hat{H}_d\,+\mu\,\hat{H}_u\,\hat{H}_d\, \\
 & \, \, 
+Y^{ij}_{\nu}\,\hat{\nu}^c_i\,\hat{L}_j\,\hat{H}_u\,
- \mu' \,\hat{\eta}\,\hat{\bar{\eta}}\,
+Y^{ij}_x\,\hat{\nu}^c_i\,\hat{\eta}\,\hat{\nu}^c_j\,
\label{eq:superpot},
\end{align} 
with the corresponding soft SUSY-breaking terms.



We denote the \vevs for  the bilepton fields as $v_{\eta}$ and $v_{\bar{\eta}}$,  for the sneutrinos of the
 \SUL doublets $\tilde L_{i}$ by $v_{L,i}$ and those of the \SUL
 singlet sneutrinos $\tilde \nu^c_{i}$ by $v_{R,i}$, with $i=1, 2, 3$.
\subsection{Results}
Two scans were performed for this model. The first scan, which we refer to as the
 ``democratic'' scan, took random values for each diagonal entry of the
 \rsnu\ -- bilepton Yukawa coupling $Y_{x}$ independently over its range. The
 other,
 which we refer to as the ``hierarchical'' scan, kept the $( 1, 1 )$ and
 $( 2, 2 )$ entries as $10^{-3}$ and $10^{-2}$ respectively.
We classify the stability results in 3 categories. The ``RPC'' category includes points for which
the input \RPCg minimum is the global minimum, the ``RPV'' category includes points for which a \RPVg minimum 
was found to be the deepest minimum. The ``unbroken'' category includes points that broke
 \SUL without breaking \UBL. Not all parameter points that are ``RPC'' at the
 one-loop level were ``RPC'' at tree level, and likewise for the ``RPV''
 category. We present a summary of the results In table \ref{tab:category_populations}.  

\begin{table} [!htbp]
\centering
\begin{tabular}{|c|c|c|c|c|} 
\hline
Categorization & \multicolumn{2}{c|}{Hierarchical scan}
                 & \multicolumn{2}{c|}{Democratic scan} \\
\hline
total & \multicolumn{2}{c|}{1640} & \multicolumn{2}{c|}{2330}\\
\hline
 & tree level & one-loop level & tree level & one-loop level \\
\hline
``RPC''      & 1422 & 1275 & 2236 & 2167 \\
\hline
``RPV''      &  218 &  212 &  94 &  86 \\
\hline
``unbroken'' &    0 &  153 &    0 &  77 \\
\hline
\end{tabular} 
\caption{Number of parameter points in the various categories.}
\label{tab:category_populations}
\end{table}

%

In fig. \ref{fig:hier_loop_comparisons} , we present these results plotted in some cuts of parameter space, including as well the result from the tunneling time calculation. There it is evident that smaller (and negative) sneutrino soft SUSY-breaking masses-squared are more likely to lead to smaller sneutrino masses and trigger \vevs for the sneutrinos, breaking $R$ parity. Also higher values for $Y_X$ play an important role, as $R$ parity is violated more often for higher values of $\mbox{Tr}[Y_X^2]$. This is because the potential has destabilizing terms such as $-v_R^2 v_{\bar{\eta}} Y_X \mu'$. However, we would like to
 point out that these are only trends and for example, \RPCg points are found for negative soft SUSY-breaking sneutrino masses-squared and vice versa, which slightly disagrees with previous results in literature \cite{FileviezPerez:2010ek}.

\section{Conclusions}
When dealing with models that have extra scalar fields and extended Higgs sectors, understanding the vacuum structure becomes crucial for phenomenological studies. 
The common approach of choosing free parameters to satisfy the minimization conditions for a desired set of \vevs leaves out a plethora of minima that can be unphysical and lower than the desired one. 
The power of the homotopy continuation method for finding tree-level
 minima combined with gradient-based minimization with loop corrections has been
 combined in the publicly-available code \vcs  \cite{CamargoMolina:2013qva}. We used this tool to
 investigate the vacuum structure of two interesting supersymmetric models and presented some of 
 the results. 
 We showed that finding the global minimum of the one-loop effective potential sheds some light in phenomenological issues like spontaneous breaking of $R$ parity in the BLSSM and the appearance of charge- and color- breaking minima in the CMSSM. The current implementation of this procedure allows for the quick analysis of broad parameter scans, therefore making it feasible to be included as an extra constraint for the exclusion of regions in parameter space.

\bibliography{SUSYstability}
\bibliographystyle{JHEP}

\end{document}